\documentclass{llncs}
\usepackage{epsfig}
\usepackage[english]{babel}
\usepackage{graphicx}
\usepackage{amssymb}
\usepackage{subfigure}
\usepackage{rotating}
\usepackage[ruled, noline, algo2e, noend]{algorithm2e}

\SetArgSty{}  \SetKw{KwOr}{or}
\SetKw{KwAnd}{and} \SetKw{KwInvariant}{invariant:}
\SetKw{KwDownTo}{down to} \SetKwInOut{Input}{input}
\SetKwInOut{Output}{output} \SetKwInOut{Require}{require}
\SetKwComment{Comment}{start}{end}
\SetKwIF{If}{ElseIf}{Else}{if}{then}{else if}{else}{endif}

\newcommand{\eat}[1] {{}}

\newtheorem{corol}{Corollary}
\newtheorem{statement}{Statement}

\begin{document}
\parindent=0cm
\parskip=2pt

\fontsize{10.5pt}{12pt}\selectfont

\title{Upward Point-Set Embeddability}
\author{
Markus Geyer, \inst1 Michael Kaufmann, \inst1 Tamara  Mchedlidze,
\inst2 Antonios Symvonis \inst2 }

\institute{
Wilhelm-Schickard-Institut f\"{u}r Informatik, Universit\"{a}t T\"{u}bingen, T\"{u}bingen, Germany.\\
\texttt{\{geyer,mk\}@informatik.uni-tuebingen.de} \and Dept. of
Mathematics, National Technical University of Athens,
 Athens, Greece.\\
    \texttt{\{mchet,symvonis\}@math.ntua.gr}
}

\maketitle

\begin{abstract}
We study the problem of Upward Point-Set Embeddability, that is the
problem of deciding whether a given upward planar digraph $D$ has an
upward planar embedding into a point set $S$. We show that any
switch tree admits an upward planar straight-line embedding into any
convex point set. For the class of $k$-switch trees, that is a
generalization of switch trees (according to this definition a
switch tree is a $1$-switch tree), we show that not every $k$-switch
tree admits an upward planar straight-line embedding into any convex
point set, for any $k \geq 2$. Finally we show that the problem of
Upward Point-Set Embeddability is NP-complete.
\end{abstract}

\section{Introduction }

A \emph{planar straight-line embedding} of a graph $G$ into a point
set $S$ is a mapping of each vertex of $G$ to a distinct point of
$S$ and of each edge of $G$ to the straight-line segment between the
corresponding end-points so that no two edges cross each other.
Gritzmann \emph{et al.}~\cite{GritzmannMPP91} proved that
outerplanar graphs is the class of graphs that admit a planar
straight-line embedding into every point set in general position or
in convex position. Efficient algorithms are known to embed
outerplanar graphs~\cite{comgeo/Bose02} and
trees~\cite{jgaa/BoseMS97} into any point set in general or in
convex position. From the negative point of view,
Cabello~\cite{Cabello06} proved that the problem of deciding whether
there exists a planar straight-line embedding of a given graph $G$
into a point set $P$  is NP-hard even when $G$ is $2$-connected and
$2$-outerplanar. For upward planar digraphs, the problem of
constructing upward planar straight-line embeddings into point sets
was studied by Giordano et al.~\cite{GiordanoLMS07}, later on  by
Binucci et al.~\cite{BinucciGDEFKL10} and recently by Angelini et
al.~\cite{AngeliniFGKMS10}. While some positive and negative results
are known for the case of upward planar digraphs, the complexity of
testing upward planar straight-line embeddability into point sets
has not been known.

In this paper  we continue the study of the problem of upward planar
straight-line embedding of directed graphs into a given point set.
Our results include:
\begin{itemize}
\item We extend the positive
results given in~\cite{AngeliniFGKMS10,BinucciGDEFKL10} by showing
that any directed switch tree admits an upward planar straight-line
embedding into every point set in convex position.
\item We study directed $k$-switch trees, a generalization of switch trees (a $1$-switch tree is exactly a switch tree).
From the construction given in~\cite{BinucciGDEFKL10} (Theorem~5),
we know that for $k\geq 4$ not every $k$-switch tree admits an
upward planar straight-line embedding into any convex point set.
Then we fill the gap for $2$ and $3$-switch trees, by showing that,
for any $k\geq 2$ there is a class of $k$-switch trees
$\mathcal{T}_n^k$, and a point set $S$ in convex position, such that
any $T\in \mathcal{T}_n^k$ does not admit an upward planar
straight-line embedding into $S$.
\item We study the computational complexity of the upward embeddability problem. More
specifically, given a $n$ vertex upward planar digraph $G$ and a set
of $n$ points on the plane $S$, we show that deciding whether there
exists an upward planar straight-line embedding of $G$ so that its
vertices  are mapped to the points of $S$ is NP-Complete. The
decision problem remains NP-Complete  even when $G$ has a single
source and the longest simple cycle of $G$ has length four and,
moreover, $S$ is an $m$-convex point set, for some integer $m>0$.
\end{itemize}

\section{Preliminaries} \label{se:definitions}
We mostly follow the terminology of~\cite{BinucciGDEFKL10}. Next, we
give some definitions that are  used throughout  this paper.

Let $l$ be a line on the plane, which is not parallel to the
$x$-axis. We say that point $p$ \emph{lies to the right of $l$}
(resp., \emph{to the left of $l$}) if $p$ lies on a semi-line that
originates on $l$, is parallel with the $x$-axis and is directed
towards $+ \infty$ (resp., $- \infty$). Similarly, if $l$ is a line
on the plane, which is not parallel to the $y$-axis, we say that
point $p$ \emph{lies above $l$} (resp., \emph{below $l$}) if $p$
lies on a semi-line that originates on $l$, is parallel with the
$y$-axis and is directed towards $+ \infty$ (resp., $- \infty$).

A \emph{point set in general position}, or \emph{general point set},
is a point set such that no three points lie on the same line and no
two points have the same $y$-coordinate. The \emph{convex hull}
$H(S)$ of a point set $S$ is the point set that can be obtained as a
convex combination of the points of $S$. A \emph{point set in convex
position}, or \emph{convex point set}, is a point set such that no
point is in the convex hull of the others. Given a point set $S$, we
denote by $b(S)$ and by $t(S)$ the lowest and the highest point of
$S$, respectively. A \emph{one-sided convex point set} $S$ is a
convex point set in which $b(S)$ and $t(S)$ are adjacent in the
border of $H(S)$.  A convex point set which is not one-sided, is
called a \emph{two-sided convex point set}. In a convex point set
$S$, the subset of points that lie to the left (resp. right) of the
line through $b(S)$ and $t(S)$ is called the \emph{left (resp.
right) part of $S$}.  A one-sided convex point set $S$ is called
\emph{left-heavy} (resp., \emph{right-heavy})convex point set if all
the points of $S$ lie to the left (resp., to the right) of the line
through $b(S)$ and $t(S)$. Note that, a one-sided convex point set
is either a left-heavy or a right-heavy convex point set.

Consider a point set $S$ and its convex hull $H(S)$. Let $S_1=S
\setminus H(S)$, $S_2 = S_1 \setminus H(S_1)$, \dots, $S_{m} =
S_{m-1} \setminus H(S_{m-1})$. If $m$ is the smallest integer such
that $S_m = \emptyset$, we say that $S$ is an \emph{$m$-convex point
set}. A subset of points of a convex point set $S$ is called
\emph{consecutive} if its points appear consecutive as we traverse
the convex hull of $S$ in the clockwise or counterclockwise
direction.

The graphs we study in this paper are directed. By $(u,v)$ we denote
an arc directed from $u$ to $v$. A \emph{switch-tree} is a directed
tree $T$, such that, each vertex of $T$ is either a source of a
sink. Note that the longest directed path of a switch-tree has
length one\footnote{The \emph{length} of a directed path is the
number of arcs in the path.}. Based on the length of the longest
path, the class of switch trees can be generalized to that of
$k$-switch trees. A \emph{$k$-switch tree} is a directed tree, such
that its longest directed path has length $k$. According to this
definition a switch tree is a $1$-switch tree. A digraph $D$ is
called \emph{path-DAG}, if its underlying graph is a simple path. A
\emph{monotone path} $(v_1,v_2,\dots,v_k)$ is a path-DAG containing
arcs $(v_i,v_{i+1})$, $1\leq i \leq k-1$.

An \emph{upward planar directed graph} is a digraph that admits a
planar drawing where each edge is represented by a curve
monotonically increasing in the $y$-direction. An \emph{upward
straight-line embedding} (\emph{UPSE} for short) of a graph into a
point set is a mapping of each vertex to a distinct point and of
each arc to a straight-line segment between its end-points such that
no two arcs cross and each arc $(u,v)$ has $y(u)<y(v)$. The
following results were presented in~\cite{BinucciGDEFKL10} and are
used in this paper.

\begin{lemma}[Binucci at al.~\cite{BinucciGDEFKL10}]
\label{lemma:binucci} Let $T$ be an $n$-vertex tree-DAG and let $S$
be any convex point set of size $n$. Let $u$ be any vertex of $T$
and let $T_1, T_2, \dots , T_k$ be the subtrees of $T$ obtained by
removing $u$ and its incident edges from $T$. In any UPSE of~$T$
into $S$, the vertices of $T_i$ are mapped into a set of consecutive
points of $S$, for each $i = 1, 2,\dots, k$.
\end{lemma}

\begin{theorem}[Binucci at al.~\cite{BinucciGDEFKL10}]
\label{theorem:binucci} For every  odd integer $n \geq 5$, there
exists a $(3n + 1)$-vertex directed tree $T$ and a convex point set
$S$ of size $3n +1$ such that $T$ does not admit an UPSE into $S$.
\end{theorem}

\section{Embedding a switch-tree  into a point set in convex position}
\label{sec:switch_tree}

In this section we enrich the positive results presented
in~\cite{AngeliniFGKMS10,BinucciGDEFKL10} by proving that, any
 switch-tree has an UPSE into any point
set in convex position. During the execution of the algorithms,
presented in the following lemmata, which embed a tree $T$ into a
point set $S$, a \emph{free point} is a point of $S$ to which no
vertex of $T$ has been mapped yet. The following lemma treats the
simple case of a one-sided convex point set and is an immediate
consequence of a result by Heath et al.~\cite{HeathPT99}.

\begin{figure}[t]
\centering
    \includegraphics[width=0.7\textwidth]{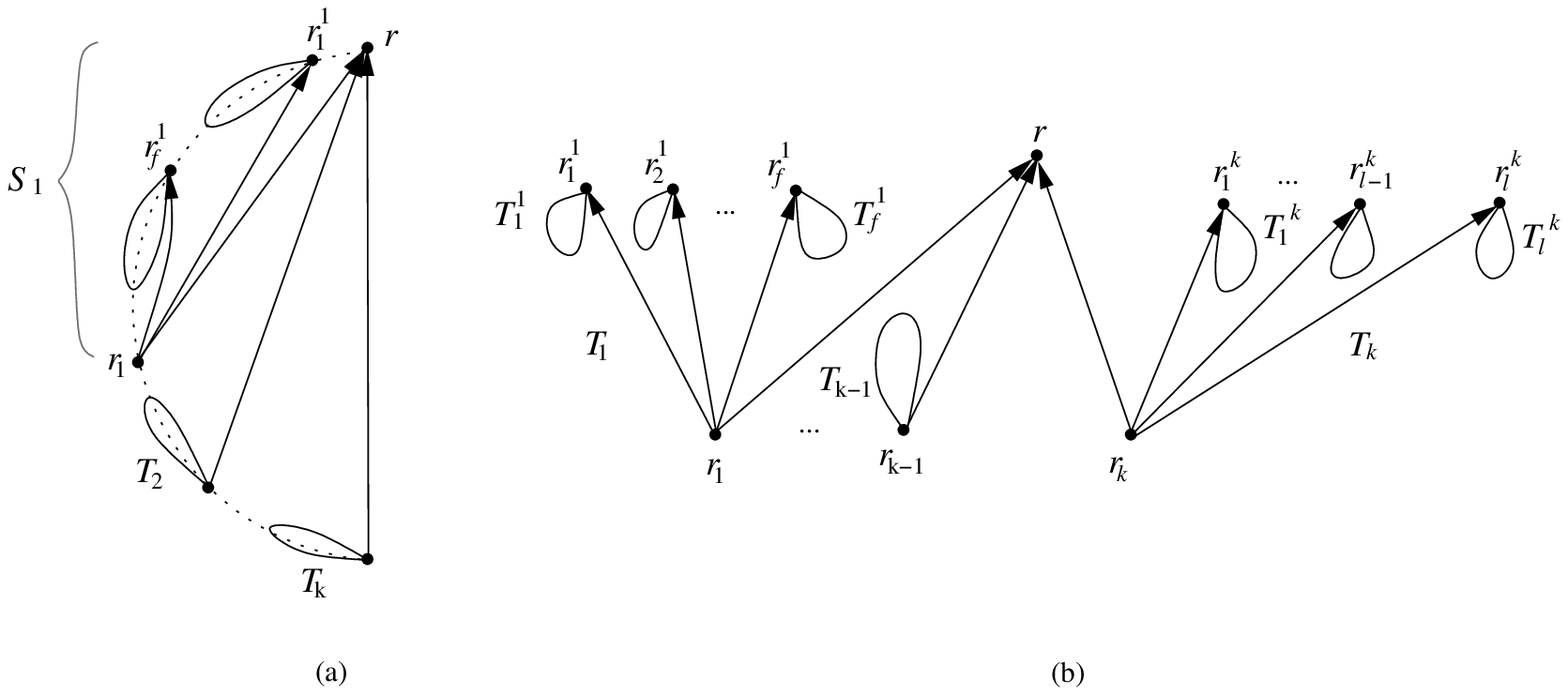}
    \caption{The construction of Lemma~\ref{lemma:stwith_tree_one_sided} and Lemma~\ref{lemma:stwith_tree_bottom}}
    \label{fig:switch_tree}
\end{figure}

\begin{lemma}
\label{lemma:stwith_tree_one_sided} Let $T$ be a switch-tree, $r$ be
a sink of $T$, $S$ be a one-sided convex point set so that
$|S|=|T|$, and $p$ be $S$'s highest point. Then, $T$ admits an UPSE
into $S$ so that vertex $r$ is mapped to point $p$.
\end{lemma}

\begin{proof}
Let $T_1, \dots, T_k$ be the sub-trees of $T$ that are connected to
$r$ by an arc and let $r_1,\dots,r_k$ be the vertices of $T_1,
\dots, T_k$, respectively, that are connected to $r$ (see
Figure~\ref{fig:switch_tree}.b). Observe that, since $T$ is a switch
tree and $r$ is a sink, vertices $r_1,\dots,r_k$ are sources. We
draw $T$ as follows: We map $r$ to $p$, then we map $T_1$ to the
$|T_1|$ highest points of $S$, so that $r_1$ is mapped to the lowest
of them. This can be trivially done if $T_1$ consists of a single
vertex, i.e. of $r_1$. Assume now that $T_1$ contains more than one
vertex. Denote by $S_1$ the $|T_1|$ highest free points of $S$. Let,
$T^1_1,\dots,T^1_f$ be the sub-trees of $T_1$, connected to $r_1$ by
an arc, and let $r^1_1,\dots,r^1_f$ be the vertices of
$T^1_1,\dots,T^1_f$, respectively, to which $r_1$ is connected.
Since $r_1$ is a source, $r^1_1,\dots,r^1_f$ are all sinks. Using
the lemma recursively we draw $T^1_1$ on the $|T^1_1|$ consecutive
highest points of $S_1$ so that $r^1_1$ is mapped to the highest
point (Figure~\ref{fig:switch_tree}.a). Similarly we draw trees
$T^1_2,\dots,T^1_f$ on the remaining free points of $S_1$. Finally
we map $r_1$ to the last free point of $S_1$, i.e. to its lowest
point. Since all of $r^1_1,\dots,r^1_f$ are drawn higher than $r_1$,
the arcs $(r_1,r^1_1),\dots,(r_1,r^1_f)$ are drawn in upward
fashion. Since each of $T^1_1,\dots,T^1_f$ is drawn on the
consecutive points of $S$ in an upward planar fashion we infer that
the drawing of $T_1$ is upward planar and is placed on the
consecutive points of $S$.

In a similar way, we map $T_2$ to the $|T_2|$ highest consecutive
free points of $S$ so that $r_2$ is mapped to the lowest of them. We
continue mapping the rest of the trees in the same way on the
remaining free points. Note that for any $i=1,\dots,k$, arc
$(r_i,r)$ does not intersect any of $H(P_j)$, where $P_j$ is a point
set where the vertices of the subtree $T_j$ are mapped. Hence for
any $i=1,\dots,k$, arc $(r_i,r)$ does not cross any other arc of the
drawing. Since $p$ is the highest point of $S$ and $r$ is mapped to
$p$, we infer that arcs $(r_i,r)$, $i=1,\dots,k$ are drawn in upward
fashion.  Since, by construction, the drawings of $T_1,\dots,T_k$
are upward and planar, we infer that the resulting drawing of $T$ is
upward and planar. \qed
\end{proof}

The following lemma is symmetrical to
Lemma~\ref{lemma:stwith_tree_one_sided} and can be proved by a
symmetric construction.

\begin{lemma}
\label{lemma:stwith_tree_one_sided_source} Let $T$ be a switch-tree,
$r$ be a source of $T$, $S$  be a one-sided convex point set so that
$|S|=|T|$, and $p$ be $S$'s lowest point. Then, $T$ admits an UPSE
into $S$ so that vertex $r$ is mapped to point $p$. \qed
\end{lemma}

\begin{figure}[t]
\centering
    \includegraphics[width=0.6\textwidth]{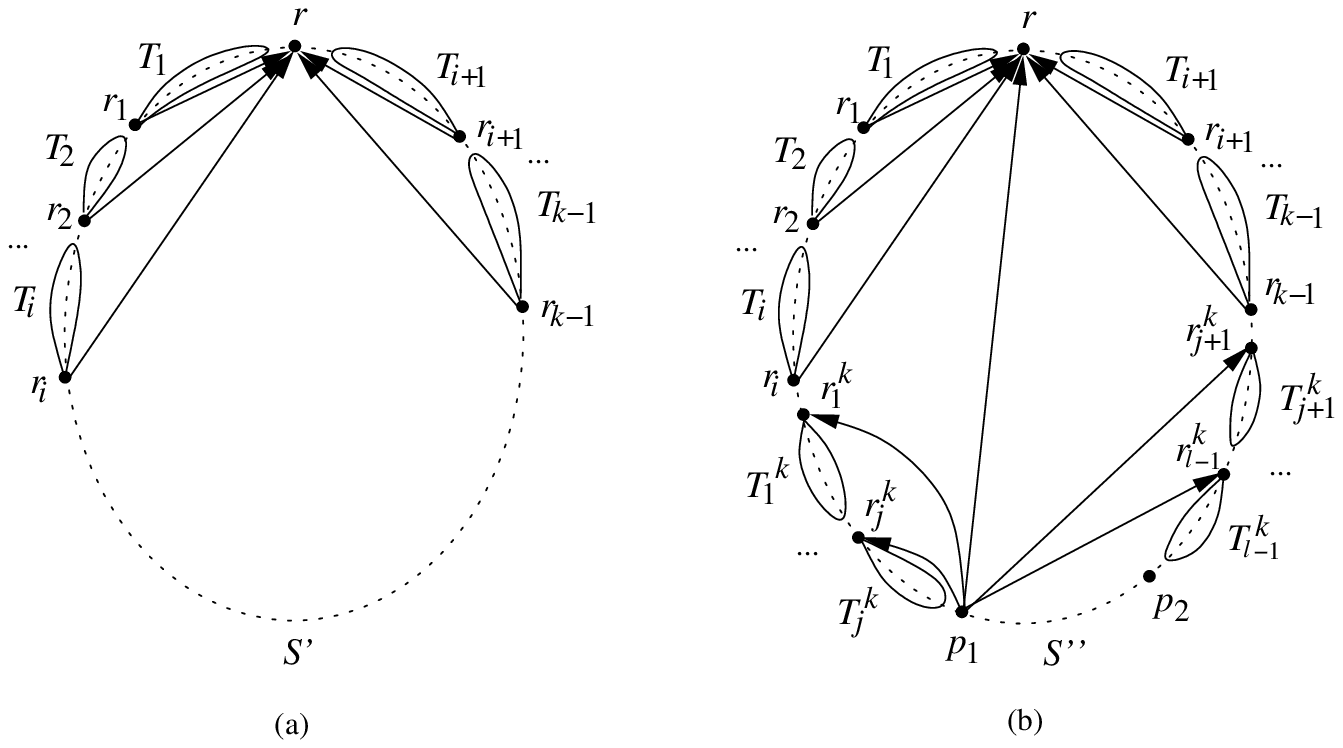}
    \caption{The construction of Lemma~\ref{lemma:stwith_tree_bottom}}
    \label{fig:switch_tree_pointset}
\end{figure}

Now we are ready to proceed to the main result of the section.

\begin{theorem}
\label{theorem:stwith_tree_bottom} Let $T$ be a switch-tree and $S$
be a convex point set such that $|S|=|T|$. Then, $T$ admits an UPSE
into $S$.
\end{theorem}

The proof of the theorem is based on the following lemma, which
extends Lemma~\ref{lemma:stwith_tree_one_sided} from one-sided
convex point sets to convex point sets.

\begin{lemma}
\label{lemma:stwith_tree_bottom} Let $T$ be a switch-tree, $r$ be a
sink of $T$, $S$ be a convex point set such that $|S|=|T|$. Then,
$T$ admits an UPSE into $S$ so that vertex $r$ is mapped to the
highest point of $S$.
\end{lemma}
\begin{proof}
Let $T_1, \dots, T_k$ be the sub-trees of $T$ that are connected to
$r$ by an edge (Figure~\ref{fig:switch_tree}.b) and let
$r_1,\dots,r_k$ be the vertices of $T_1, \dots, T_k$, respectively,
that are connected to $r$. Observe that, since $T$ is a switch tree
and $r$ is a sink, vertices $r_1,\dots,r_k$ are sources.

We draw $T$ on $S$ as follows. We start by placing the trees
$T_1,T_2,\dots$ on the left side of the point set $S$ as long as
they fit, using the highest free points first. This can be done in
an upward planar fashion by
Lemma~\ref{lemma:stwith_tree_one_sided_source}
(Figure~\ref{fig:switch_tree_pointset}.a).
   Assume that $T_{i}$ is the last placed subtree. Then, we continue placing the trees $T_{i+1},
   \dots,T_{k-1}$ on
   the right side of the point set $S$.
   This can be done due to Lemma~\ref{lemma:stwith_tree_one_sided_source}.
   Note that the remaining free points are consecutive point of $S$, denote these points by $S^\prime$.
   To complete the embedding we draw $T_k$ on $S^\prime$. Let $T_1^k,\dots,T_l^k$ the subtrees of
$T_k$, that are connected to $r_k$ by an arc. Let also
   $r_1^k,\dots,r_l^k$ be the vertices of $T_1^k,\dots,T_l^k$, respectively, that are connected to $r_k$ (Figure~\ref{fig:switch_tree}.b). Note that
   $r_1^k,\dots,r_l^k$ are all sinks. We start by drawing
   $T_1^k,T_2^k,\dots$ as long
   as they fit on the left side of point set $S^\prime$, using the highest free points first.  This can be done in an upward planar fashion by Lemma~\ref{lemma:stwith_tree_one_sided}.
   Assume that $T_j^k$ is the last placed subtree (Figure~\ref{fig:switch_tree_pointset}.b). Then, we continue on
   the right side of the point set $S^\prime$ with the trees $T_{j+1}^k,\dots,T_{l-1}^k$.
   This can be done again by Lemma~\ref{lemma:stwith_tree_one_sided}.
   Note that there are exactly $|T^k_l|+1$ remaining free points
   since we have not yet drawn $T_l^k$ and vertex $r_k$ of $T_k$.
   Denote by $S^{\prime \prime}$ the remaining free points and note that $S^{\prime \prime}$ consists of consecutive points of $S$. If $S^{\prime \prime}$ is a one-sided point set
   then we can proceed by using the Lemma~\ref{lemma:stwith_tree_one_sided} again and the result follows trivially.
   Assume now that $S^{\prime \prime}$ is a two-sided convex point set and let $p_1$ and $p_2$
   be the highest  points of $S^{\prime \prime}$ on the left and on the right, respectively.
   W.l.o.g., let $y(p_1) < y(p_2)$. Then, we map $r_k$
   to $p_1$. By using the lemma recursively, we can draw $T^k_l$ on
   $S^{\prime \prime} \setminus \{p_1\}$ so that $r_l^k$ is
   mapped to $p_2$. The proof is completed by observing that
   all edges connecting $r_k$ to $r_1^k,\dots,r_l^k$
   and $r_1,\dots,r_k$ to $r$ are upward and do not cross each other.
   \qed
\end{proof}

Theorem~\ref{theorem:stwith_tree_bottom} follows immediately if we
select any sink-vertex of $T$ as $r$ and apply
Lemma~\ref{lemma:stwith_tree_bottom}.

\section{$K$-switch trees} \label{sec:k_switch_tree}

Binucci et al.~\cite{BinucciGDEFKL10} (see also
Theorem~\ref{theorem:binucci}) presented a class of trees and
corresponding convex point sets, such that any tree of this class
does not admit an UPSE into its corresponding point set.

\begin{figure}[t]
\centering
    \includegraphics[width=0.6\textwidth]{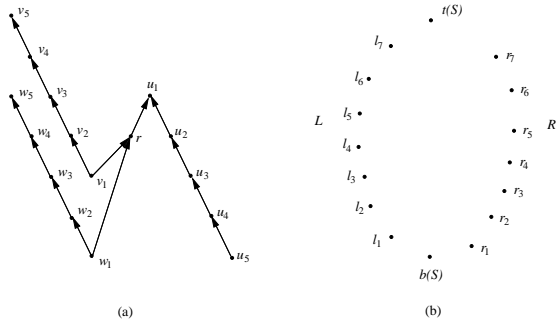}
    \caption{(a-b) A $4$-switch tree $T$ and a point set $S$, such that $T$ does not admit an UPSE into point set $S$.}
\label{fig:4_switch_tree}
\end{figure}

The $(3n+1)$-size tree $T$ constructed in the proof of
Theorem~\ref{theorem:binucci}\cite{BinucciGDEFKL10} has the
following structure (see Figure~\ref{fig:4_switch_tree}.a for the
case $n=5$). It consists of: ($i$) one vertex $r$ of degree three,
($ii$) three monotone paths of $n$ vertices: $P_u = (u_n, u_{n-1},
\dots , u_1)$, $P_v = (v_1, v_2,\dots, v_n)$, $P_w = (w_1, w_2,
\dots, w_n)$, ($iii$) arcs $(r, u_1)$, $(v_1,r)$ and $(w_1,r)$.

The $(3n + 1)$-convex point set $S$, used in the proof of
Theorem~\ref{theorem:binucci}\cite{BinucciGDEFKL10}, consists of two
extremal points on the $y$-direction, $b(S)$ and $t(S)$, the set $L$
of $(3n - 1)/2$ points $l_1, l_2, \dots, l_{(3n-1)/2}$, comprising
the left side of $S$ and the set $R$ of $(3n - 1)/2$ points $r_1,
r_2, \dots, r_{(3n-1)/2}$, comprising the right side of $S$. The
points of $L$ and $R$ are located so that $y(b(S)) < y(r_1) < y(l_1)
< y(r_2) < y(l_2) < \dots < y(r_{(3n-1)/2}) < y(l_{(3n-1)/2}) <
y(t(S))$. See Figure~\ref{fig:4_switch_tree}.b for $n=5$.

Note that the $(3n+1)$-node tree $T$ described above is a
$(n-1)$-switch tree. Hence a straightforward corollary of
Theorem~\ref{theorem:binucci}\cite{BinucciGDEFKL10} is the following
statement.

\begin{corol}
\label{stat:k-switch} For any $k\geq 4$, there exists a $k$-switch
tree $T$ and a convex point set $S$ of the same size, such that $T$
does not admit an UPSE into $S$.
\end{corol}

From Section~\ref{sec:switch_tree}, we know that any switch tree
$T$, i.e. a $1$-switch tree, admits an UPSE into any convex point
set. The natural question raised by this result and
Corollary~\ref{stat:k-switch} is whether an arbitrary $2$-switch or
$3$-switch tree has an UPSE into any convex point set. This question
is resolved by the following theorem.

\eat{
\begin{theorem}
\label{theorem:2_3_switch} There exists a $2$-switch tree $T_2$, a
$3$-switch tree $T_3$, and a convex point set $S$ of the same size,
such that neither $T_2$ nor $T_3$ admits an UPSE into point set $S$.
\end{theorem}

\begin{figure}[tbh]
\centering
    \includegraphics[width=0.8\textwidth]{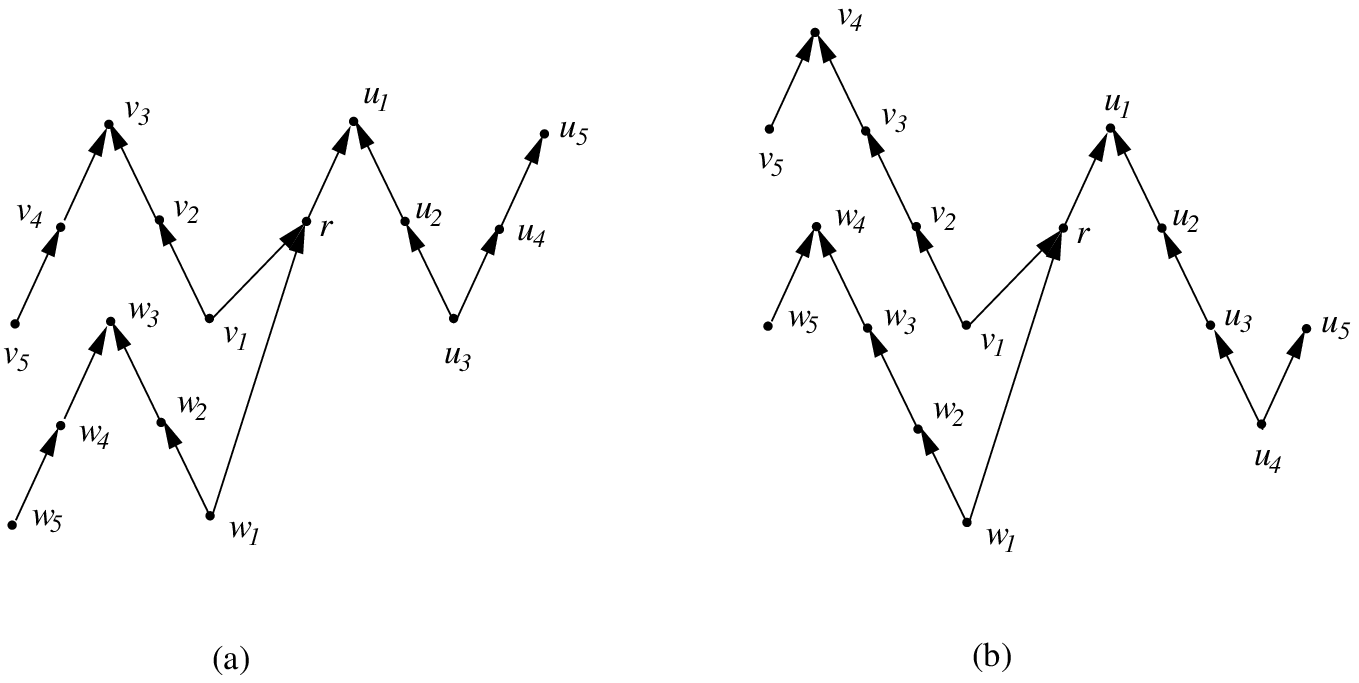}
    \caption{(a) A $2$-switch tree $T_2$. (b) A $3$-switch tree
    $T_3$.}
    \label{fig:3_switch_tree}
\end{figure}

\begin{proof}
Consider the trees $T_2$ and $T_3$ of
Figures~\ref{fig:3_switch_tree}.a,b respectively and a point set $S$
of Figure~\ref{fig:4_switch_tree}.b. We will show that neither $T_2$
nor $T_3$ has an UPSE into $S$. Note that the trees $T_2$ and $T_3$
has their underling graph identical to the underling graph of the
$4$-switch tree $T$, used by Binucci et al~\cite{BinucciGDEFKL10},
while the point set used in our proof is identical to that used by
authors of~\cite{BinucciGDEFKL10}. The proof is identical for both
trees $T_2$ and $T_3$.

For the sake of contradiction we assume that there exists an UPSE of
$T_2$ (resp. $T_3$) into $S$. Let $P_u$, $P_v$, $P_w$ be the
subtrees of $T_2$ (resp. $T_3$) obtained by removing $r$ and its
incident edges from $T$, so that $V(P_u)=\{u_1,\dots,u_5\}$,
$V(P_v)=\{v_1,\dots,v_5\}$ and $V(P_w)=\{w_1,\dots,w_5\}$. By
Lemma~\ref{lemma:binucci} each of $P_u$, $P_v$ and $P_w$ is drawn on
the consecutive points of $S$. Call $S_u$, $S_v$ and $S_w$ the
subsets of point set $S$, where $P_u$, $P_v$ and $P_w$ are mapped
to, respectively.   Consider the largest subset $S^\prime$ of the
point set $S$ that is a one-sided convex point set. Note that
$|S^\prime|=9$ and hence at least one of $S_u$, $S_v$, $S_w$ is a
two-sided point set. Next we consider two cases based on whether
$S_u$ or $S_v$ is a two-sided convex point set. The case when $S_w$
is a two-sided convex point set is similar to that of $S_v$.

\begin{figure}[tbh]
\centering
    \includegraphics[width=0.6\textwidth]{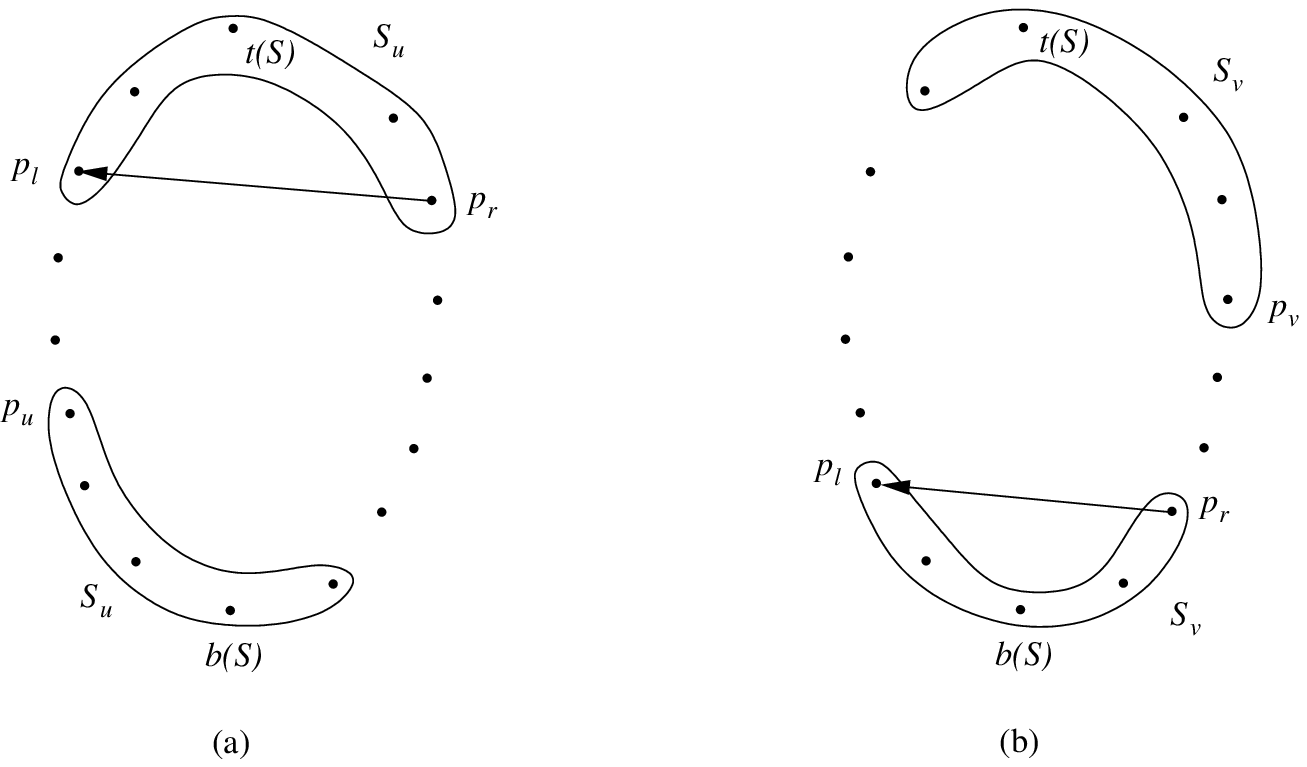}
    \caption{(a-b) The construction used in the proof of Theorem~\ref{theorem:2_3_switch}.}
    \label{fig:2_3_switch_tree_proof}
\end{figure}

\textbf{Case~$1$:} $S_u$ is a two-sided convex point set. Assume
first that $S_u$ contains $b(S)$
(Figure~\ref{fig:2_3_switch_tree_proof}.a). Denote by $p_u$ the
point of $S_u$ with the largest $y$-coordinate. Note that
$S\setminus S_u$ contains at most two points lower than $p_u$. Note
also that $r,~v_1,~w_1$ have to be placed lower than $p_u$. Hence
the drawing can not be upward. A clear contradiction. Let now, that
$S_u$ contain $t(S)$. Denote by $p_l$ and $p_r$ the lowest points of
the left and of the right side of $S_u$. Note that any UPSE of $P_u$
on $S_u$ (either for $T_2$ or for $T_3$) contains an edge connecting
point $p_l$ and $p_r$. Also note that $u_1$ can not be mapped
neither to $p_l$ nor to $p_r$ and hence the arc $(r,u_1)$ crosses
the arc connecting $p_l$ and $p_r$. A clear contradiction.

\textbf{Case~$2$:} $S_v$ is a two-sided convex point set. Assume
first that $S_v$ contains $b(S)$
(Figure~\ref{fig:2_3_switch_tree_proof}.b). Denote by $p_l$ and
$p_r$ the highest points of the left and of the right side of $S_v$.
Note that any UPSE of $P_v$ on $S_v$ (either for $T_2$ or for $T_3$)
contains an edge connecting point $p_l$ and $p_r$. Also note that
$v_1$ can not be mapped to $p_l$ or to $p_r$ and hence the arc
$(v_1,r)$ crosses the arc connecting $p_l$ and $p_r$. A clear
contradiction. Let now, that $S_v$ contain $t(S)$. Denote by $p_v$
the point of $S_v$ with the smallest $y$-coordinate. Note that
$S\setminus S_v$ contains at most two points higher than $p_v$. Note
also that $r,~u_1$ have to be placed higher than $p_v$. Hence $P_u$
is not placed on the consecutive points of $S$. A clear
contradiction. \qed
\end{proof}
} 

\begin{theorem}
\label{theorem:general} For any $n\geq 5$ and for any $k\geq 2$,
there exists a class $\mathcal{T}_n^k$ of $3n+1$-vertex $k$-switch
trees and a convex point set $S$, consisting of $3n+1$ points, such
that any $T \in \mathcal{T}_n^k$ does not admit an UPSE into $S$.
\end{theorem}
\begin{proof}
For any $n\geq 5$ we construct the following class of trees (see
Figure~\ref{fig:k_switch_tree}.a). Let $P_u$ be an $n$-vertex
path-DAG on the vertex set $\{u_1,u_2,\dots,u_n\}$, enumerated in
the order they are presented in the underlying undirected path of
$P_u$, and such that arcs $(u_3,u_2),~(u_2,u_1)$ are present in
$P_u$. Let also $P_v$ and $P_w$ be two n-vertex path-DAGs on the
vertex sets $\{v_1,v_2,\dots,v_n\}$ and $\{w_1,w_2,\dots,w_n\}$
respectively, enumerated in the order they are presented in the
underlying undirected path of $P_v$ and $P_w$, and such that arcs
$(v_1,v_2),(v_2,v_3)$ and $(w_1,w_2),(w_2,w_3)$ are present in $P_v$
and $P_w$, respectively. Let $T(P_u, P_v, P_w)$ be a tree consisting
of $P_u$, $P_v$, $P_w$, vertex $r$ and arcs
$(r,u_1),(v_1,r),(w_1,r)$.

Let $\mathcal{T}_n^k =\{T(P_u, P_v, P_w)~|~\textrm{the longest
directed path in } P_u, P_v \textrm{ and } P_w \textrm{ has length }
k\}$, $k \geq 2$. So, $\mathcal{T}_n^k$ is a class of $3n+1$-vertex
$k$-switch trees. Let $S$ be a convex point set as described in the
beginning of the section. Next we show that any $T \in
\mathcal{T}_n^k$ does not admit an UPSE into point set $S$.

\begin{figure}[t]
\centering
    \includegraphics[width=\textwidth]{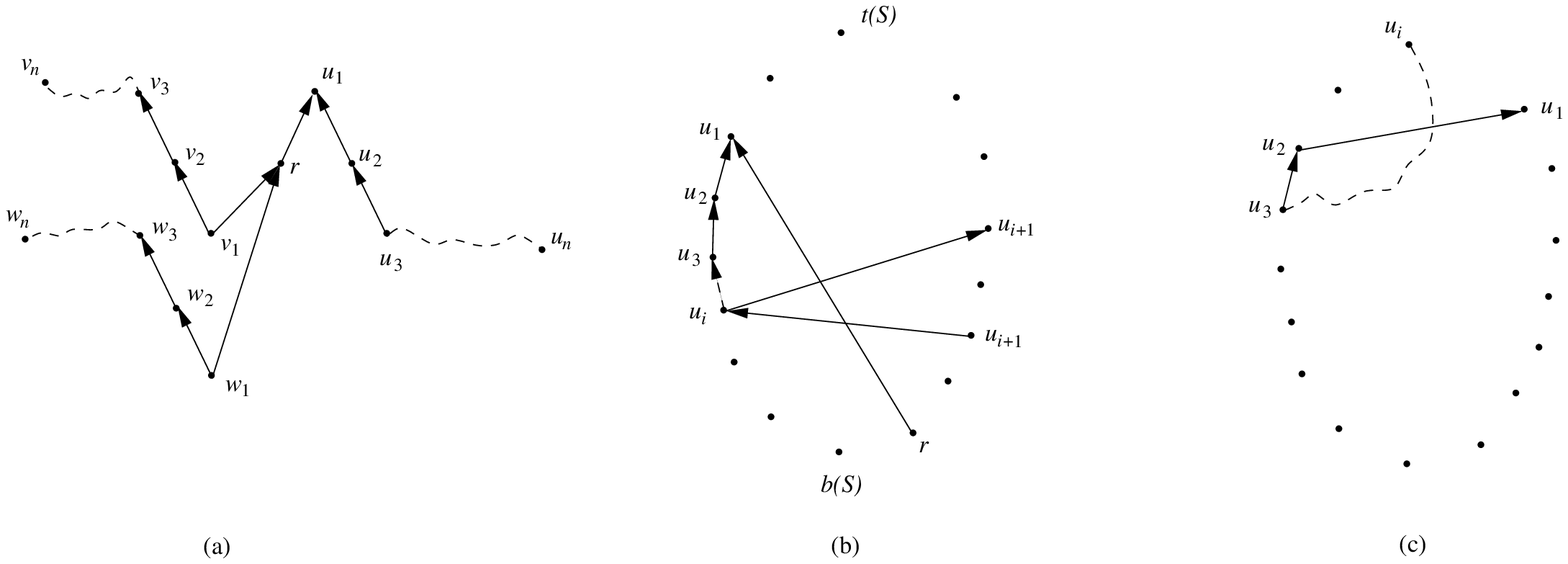}
    \caption{(a) $k$-switch tree, $k\geq 2$. (b) The construction of the proof of Statement~\ref{stat:P_u_vs_S_t}, Cases~1 to~2.}
    \label{fig:k_switch_tree}
\end{figure}

Let $T\in \mathcal{T}_n^k$. For the sake of contradiction, we assume
that there exists an UPSE of $T$
 into $S$.  By Lemma~\ref{lemma:binucci}, each of the paths $P_u$, $P_v$ and $P_w$ of $T$ is drawn on
 consecutive points of $S$. Denote by $S_u$, $S_v$ and $S_w$ the subsets
of point set $S$, in which $P_u$, $P_v$ and $P_w$ are mapped to,
respectively. Hence $|S_u|=|S_v|=|S_w|=n$. By construction of $S$,
the largest subset of $S$ which is a one-sided convex point set,
contains two extremal points of $S$ and has size
$\lceil\frac{3n-1}{2}\rceil +2 < 2n$, when $n\geq 5$. Thus, at least
one of $S_u$, $S_v$ and $S_w$ is a two-sided convex point set. We
denote by $S_b$ and $S_t$ any two-sided point sets, which consist of
consecutive points  of $S$, so that $|S_b|=|S_t|=n$, and $b(S)\in
S_b$, $t(S) \in S_t$ respectively. Next, we show that in any UPSE of
$T$ on $S$, $P_u$ can not be drawn on $S_b$, while $P_v$ and $P_w$
can not be drawn on $S_t$.

\begin{statement}
\label{stat:P_u_vs_S_t} For any upward drawing of $P_u$ on $S_t$
there is a crossing created by the arcs of $T$.
\end{statement}
\emph{Proof of Statement~\ref{stat:P_u_vs_S_t}.} Recall that $S_t
\subset S$ is a two-sided convex point set, so that $t(S) \in S_t$.
In any drawing of $P_u$ on $S_t$, the vertices $u_1,u_2,u_3$ are
mapped to some points of $S_t$. Next we consider four cases based on
whether $u_1,u_2,u_3$ are drawn on the same side of $S$.

\textbf{Case~1.} Vertices $u_1,u_2,u_3$ are mapped to the same side
of $S$, possibly including $t(S)$, say w.l.o.g. to the left side of
$S$, see Figure~\ref{fig:k_switch_tree}.b. Let $u_{i+1}$ be the
first vertex of $P_u$ that is mapped to the right side of $S$. Then,
since $r$ is mapped to a point of $S\setminus S_t$, arc $(r,u_1)$
crosses arc $(u_i,u_{i+1})$ (or arc $(u_{i+1},u_i)$).

\textbf{Case~2.} Vertices $u_2,u_3$ are mapped to the same side of
$S$, possibly including $t(S)$, say w.l.o.g. to the left side of
$S$, see Figure~\ref{fig:k_switch_tree}.c. Then, $u_1$ is mapped to
the right side of $S$. Note that $u_2$ can not be mapped to $t(S)$,
because then there is no point for $u_1$ to be mapped to, so that
the drawing is upward. Hence, there is at least one point $p$
 higher than the end points of arc $(u_2,u_1)$, that has to be visited by path $P_u$. Thus, path $P_u$
 crosses arc $(u_2,u_1)$.

\textbf{Case~3.} Vertices $u_1,u_2$ are mapped to the same side of
$S$, possibly including $t(S)$, say w.l.o.g. to the left side of
$S$. Then, $u_3$ is mapped to the right side of $S$
(Figure~\ref{fig:proof_k_switch}.a) and, as a consequence, arcs
$(r,u_1)$ and $(u_3,u_2)$ cross.

\begin{figure}[t]
\centering
    \includegraphics[width=\textwidth]{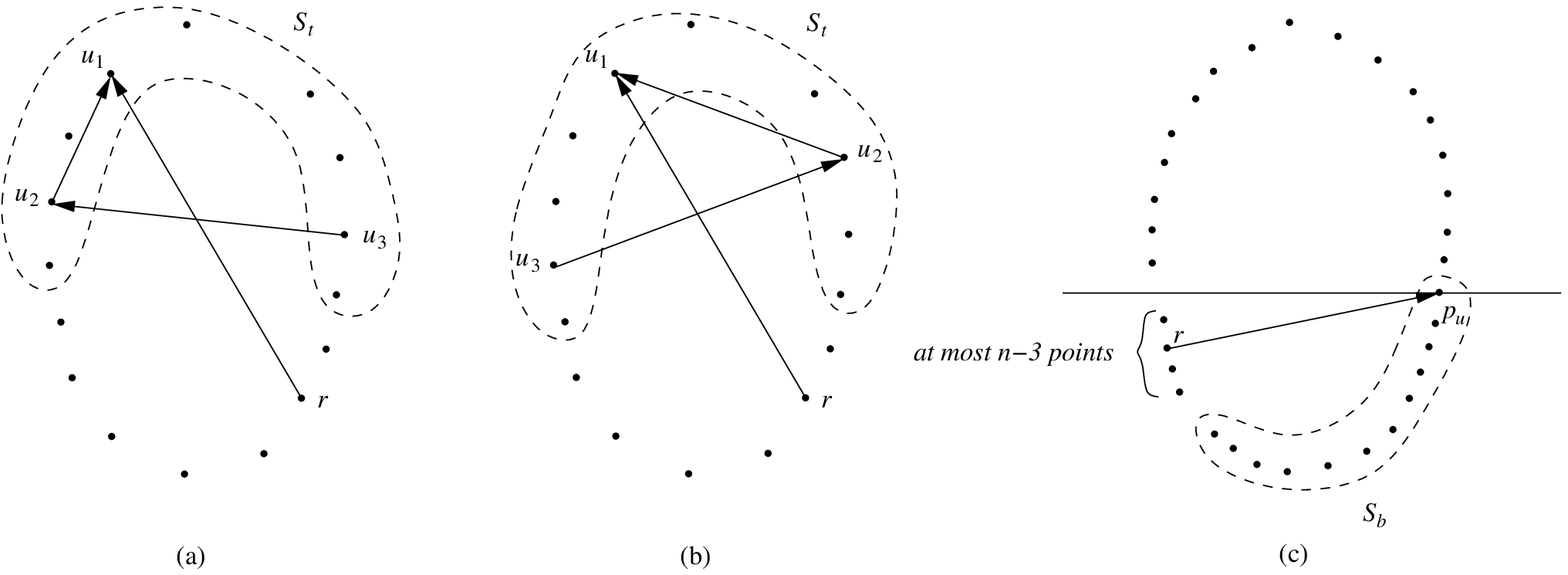}
    \caption{(a-b) The construction of the proof of Statement~\ref{stat:P_u_vs_S_t}, Cases~2 to~3. (c) The construction used in Statement~\ref{stat:P_u_vs_S_b}.}
    \label{fig:proof_k_switch}
\end{figure}

\textbf{Case~4.} Vertices $u_1,u_3$ are mapped to the same side of
$S$, possibly including $t(S)$, say w.l.o.g. to the left side of
$S$. Then, $u_2$ is mapped to the right side of $S$
(Figure~\ref{fig:proof_k_switch}.b) and, as a consequence, arcs
$(r,u_1)$ and $(u_3,u_2)$ cross. \qed

\smallskip The proof of following statement is symmetrical to the
proof of Statement~\ref{stat:P_u_vs_S_t}.

\begin{statement}
\label{stat:P_v_vs_S_b} For any upward drawing of $P_u$ or $P_w$ on
$S_b$ there is a crossing created by the arcs of $T$. \qed
\end{statement}

So, we have proved that there is no upward planar mapping of $T$
into $S$ so that $P_u$ is mapped to a set $S_t$, or such that $P_v$
or $P_w$ is mapped to a set $S_b$. Next, we prove that there is also
no upward planar mapping of $T$ on $S$ so that $P_u$ is mapped to
$S_b$, and such that $P_v$ or $P_w$ is mapped to $S_t$.

\begin{statement}
\label{stat:P_u_vs_S_b} There is no upward drawing of $T$ on point
set $S$, such that $P_u$ is mapped to the points of $S_b$.
\end{statement}
\emph{Proof of Statement~\ref{stat:P_u_vs_S_b}.} Denote by $p_u$ the
point of $S_b$ with the largest $y$-coordinate, see
Figure~\ref{fig:proof_k_switch}.c. By the construction of $S$ and
since $S_b$ is a two-sided point-set which contains $n$ points, we
infer that $S\setminus S_b$ contains at most $n-3$ points lower than
$p_u$. Moreover, all of these points are on the side opposite to
$p_u$. We observe the following: $(i)$ $r$ has to be placed lower
than $p_u$, and hence $r$ is placed on the opposite side of that of
$p_u$, $(ii)$  $v_1$ has to be placed lower than $r$, and since
$P_v$ has to be mapped to consecutive points of $S$, the whole $P_v$
is mapped to the points on the same side with $r$ and lower than
$r$. But, there are at most $n-4$ free points, a clear contradiction
since $|P_v|=n$. \qed

\smallskip The following statement is symmetrical to Statement~\ref{stat:P_u_vs_S_b}.

\begin{statement}
\label{stat:P_v_vs_S_t} There is no upward drawing of $T$ on point
set $S$, such that $P_v$ or $P_w$ is mapped to the points of $S_t$.
\qed
\end{statement}

As we observed in the beginning of the proof of the theorem, at
least one of $P_u,P_v,P_w$ is mapped to a two-sided point set
containing either $b(S)$ or $t(S)$. But, as it is proved in
Statements~\ref{stat:P_u_vs_S_t} to~\ref{stat:P_v_vs_S_t} this is
impossible. So, the theorem follows. \qed

\end{proof}

\section{Upward planar straight-line point set embeddability is NP-complete}
\label{sec:NP_complete}

In this section we examine the complexity of testing whether a given
$n$-vertex upward planar digraph $G$  admits an UPSE into a point
set $S$. We show that the problem is NP-complete even for a single
source digraph $G$ having longest simple cycle of length at most
$4$. This result is optimal for the class of cyclic
graphs\footnote{A digraph is \emph{cyclic} if its underling
undirected graph contains at least one cycle.}, since Angelini et
al.~\cite{AngeliniFGKMS10} showed that every single-source upward
planar directed graph with no cycle of length greater than three
admits an UPSE into every point set in general position.

\begin{theorem}
Given an $n$-vertex upward planar digraph $G$ and a planar point set
$S$ of size $n$ in general position, the decision problem of whether
there exists an UPSE of $G$ into $S$ is $NP$-Complete. The decision
problem remains NP-Complete even when $G$ has a single source and
the longest simple cycle of $G$ has length at most $4$ and,
moreover, $S$ is an $m$-convex point set for some $m>0$.
\end{theorem}
\begin{proof}
The problem is trivially in NP. In order to prove the
NP-completeness, we construct a reduction from the $3$-Partition
problem.

\begin{quote}
\emph{Problem: }$3$-Partition\\
\emph{ Input:} A bound $B \in \mathbb{Z^+}$, and a set
$A=\{a_1,\dots,a_{3m}\}$ with
$a_i \in \mathbb{Z^+}$, $\frac{B}{4}<a_i<\frac{B}{2}$.\\
\emph{Output:} $m$ disjoint sets $A_1,\dots,A_m \subset A$ with
$|A_i|=3$ and $\sum_{a\in A_i}{a}=B, ~1 \leq i \leq m$.
\end{quote}

We use the fact that $3$-Partition is a strongly NP-hard problem,
i.e. it is NP-hard even if $B$ is bounded by a polynomial in
$m$~\cite{GareyJ79}.
\begin{figure}[tb]
\centering
    \includegraphics[width=\textwidth]{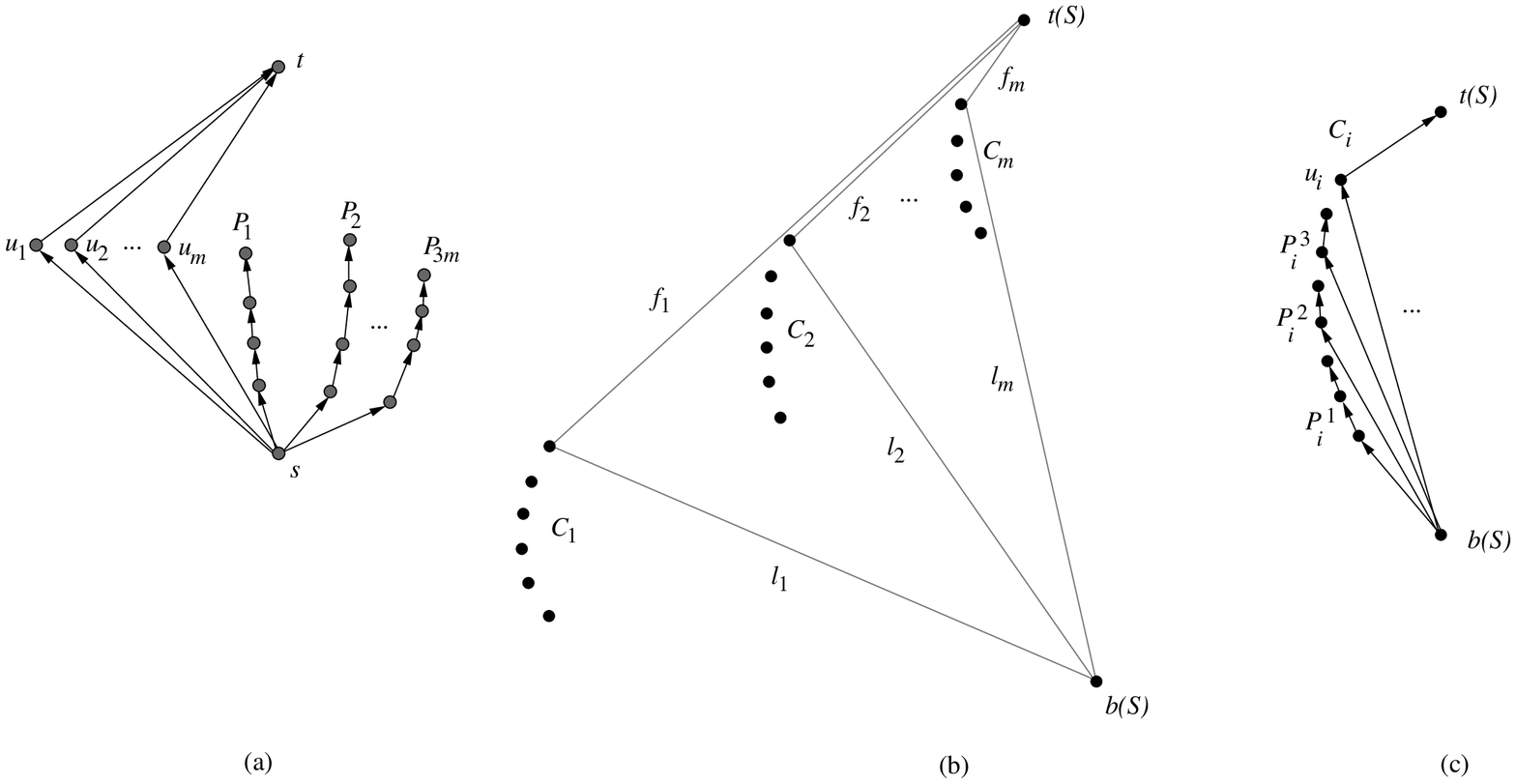}
    \caption{$(a)$ The graph $G$ of the construction used in the proof of NP-completeness.
$(b)$ The point set $S$ of the construction. $(c)$ An UPSE of $G$ on
$S$. $(d)$ The construction of Statement~\ref{stat:prop_pointset}.}
    \label{fig:construction}
\end{figure}
Let $A$ and $B$ be the set of the $3m$ positive integers  and the
bound, respectively, that form the  instance $(A, B)$ of the
$3$-Partition problem. Based on $A$ and $B$, we show how to
construct  an upward planar digraph $G$ and a point set $S$ such
that $G$ has an UPSE on point set $S$ if and only if the instance
$(A,B)$ of the 3-partition problem has a solution.

We first show how to construct $G$  (see
Figure~\ref{fig:construction}.a for illustration). We start the
construction of  $G$ by first adding  two vertices $s$ and $t$.
Vertex $s$ is the single source of the whole graph.  We then add $m$
disjoint paths from $s$ to $t$, each of length two. The degree-2
vertices of these paths are denoted by $u_i, ~i=1,\dots,m$. For each
$a \in A$, we construct a monotone directed  path $P_i$ of length
$a$ that has $a$ new vertices and   $s$ at its source. Totally, we
have $3m$ such paths $P_1,\dots,P_{3m}$.

We proceed to the construction of  point set $S$. Let $b(S)$ and
$t(S)$ be the lowest and the highest points of $S$ (see
Figure~\ref{fig:construction}.b). In addition to $b(S)$ and $t(S)$,
$S$ also contains $m$ one-sided convex point sets $C_1,\dots,C_m$,
each of size $B+1$, so that the points of $S$ satisfy the following
properties:

\begin{itemize}
\item $C_i \cup \{b(S),t(S)\}$ is a
left-heavy convex point set,  $i\in \{1,\dots,m\}$.
\item The points of $C_{i+1}$ are higher than  the
points of $C_i$, $i\in \{1,\dots,m-1\}$.
\item Let $l_i$ be the line
through $b(S)$ and $t(C_i)$, $i\in \{1,\dots,m\}$. $C_1,\dots,C_{i}$
lie to the left of line $l_i$ and  $C_{i+1},\dots,C_{m}$ lie to the
right of line $l_i$.
\item Let $f_i$ be the  line through $t(S)$ and $t(C_i)$, $i\in \{1,\dots,m\}$.
$C_j$, $j\geq i$, lie to the right of line $f_i$.
\item $\{t(C_i):i=1,\dots,m\}$ is a left-heavy convex point
set.
\end{itemize}

The next statement follows from the properties of point set $S$.

\begin{statement}
\label{stat:prop_pointset} Let $C_i$ be one of the left-heavy convex
point sets comprising $S$ and let $x \in C_j$, $j>i$. Then, set $C_i
\cup \{b(S),x\}$ is also a left-heavy convex point set, with $b(S)$
and $x$ consecutive on its convex hull. \qed
\end{statement}

\eat{ \emph{Proof of Statement~\ref{stat:prop_pointset}:} Let
$p^i_1, p^i_2$ be two highest points of $C_i$. Assume a line $l$,
though $p^i_1, p^i_2$, see Figure~\ref{fig:construction}.d. Let also
$l_t$ be a line segment through $p^i_1$ and $t(S)$. Note that the
line $l_t$ is to the right of $l$, i.e. each point of $l_t$ is to
the right of $l$, except of the point $p^i_1$ where they cross. Note
also that the point $x$, since $x \in C_j$, such that $j>i$, is to
the right of $l_t$, due to the structure of the point set $S$. I.e.
rotating clockwise the line $l$, fixed in $p^i_1$, the first point
of the set $C_i \cup \{b(S),x\}$ encountered is the point $x$. Then
note that, if we take a line $l_x$, through $x$ and $p^i_1$ and
rotate it clockwise fixed at $x$, the first point encountered if
$b(S)$. So, $C_i \cup \{b(S),x\}$ is a convex point set, moreover it
is left-heavy, since $x$ and $b(S)$ are consecutive on its convex
hull and are the lowest and the highest points of $C_i \cup
\{b(S),x\}$, respectively. \qed }

\begin{statement}
\label{statement:area} We can construct  a point set $S$ that
satisfies  all the above requirements so that the area of $S$ is
polynomial on $B$ and $m$.
\end{statement}
\emph{Proof of Statement~\ref{statement:area}:} For  each $i\in
\{0,\dots,m-1\}$ we let $C_{m-i}$ to be the set of $B+1$ points
$$C_{m-i}= \left\{ ( -j-i(B+2), j^2 -(i(B+2))^2 ) ~|~~j=1,2,\ldots, B+1 ~ \right\} $$

Then, we set the lowest point of the set $S$, called $b(S)$, to be
point $(-(B+1)^2 + ((m-1)(B+2))^2, (B+1)^2 - (m(B+2))^2)$ and the
highest point of  $S$, called $t(S)$, to be point $(0,(m(B+2))^2)$.

\eat{(see Figure~\ref{fig:pointset_matlab} for an example point
set).}

It is easy to verify that all the above requirements hold and that
the area of the rectangle bounding the constructed point set is
polynomial on $B$ and $m$. \qed

\eat{
\begin{sidewaysfigure}
    \centering
    \includegraphics[width=\textwidth]{imagesnp/point_set_2}
    \caption{A real point set constructed in Matlab for $B=3$ and $m=5$. For better visual understanding the point set is represented by
    dark lines (not points). The whole point set consists of the all the dark lines and two extremal point to which they are connected.}
    \label{fig:pointset_matlab}
\end{sidewaysfigure}
}

\begin{statement}
\label{statement:size} $|S|=|V(G)|=m(B+1)+2$. \qed
\end{statement}

We now proceed to show how from a solution for the $3$-Partition
problem we can derive a solution for the upward point set
embeddability problem. Assume that there exists a solution for the
instance of the $3$-Partition problem and let it be
$A_i=\{a_i^1,a_i^2,a_i^3\}$, $i=1 \ldots m$. Note that
$\sum_{j=1}^{3} a_i^j=B$. We first map $s$ and $t$ to $b(S)$ and
$t(S)$, respectively.  Then, we map vertex $u_i$ on $t(C_i)$, $i=1
\ldots m$. Note that the path from $s$ to $t$ through $u_i$  is
upward and $C_1,\dots,C_{i}$ lie entirely to the left of this path,
while $C_{i+1},\dots,C_m$ lie to the right of this path. Now each
$C_i$ has $B$ free points. We map the vertices of paths $P_i^1$,
$P_i^2$ and $P_i^3$ corresponding to $a_i^1,a_i^2,a_i^3$ to the
remaining points of $C_i$ in an upward fashion (see
Figure~\ref{fig:construction}.c). It is easy to verify that the
whole drawing is upward and planar.

Assume now that there is an UPSE of $G$ into $S$. We prove that
there is a solution for the corresponding $3$-Partition problem. The
proof is based on the following statements.

\begin{statement}
\label{statement:bottom} In any UPSE of $G$ into $S$, $s$ is mapped
to $b(S)$. \qed
\end{statement}

\begin{statement}
\label{statement:one_u} In any UPSE of $G$ into $S$,  only one
vertex from set $\{ u_1, \ldots u_m \}$ is mapped to
 point set  $C_i$, $i=1 \ldots m$.
\end{statement}
\emph{Proof of Statement~\ref{statement:one_u}:} For the sake of
contradiction, assume that there are two distinct vertices $u_j$ and
$u_k$ that are mapped to two points of the same point set $C_i$ (see
Figures~\ref{fig:two_u}). W.l.o.g. assume that $u_k$ is mapped to a
point higher than the point $u_j$ is mapped to. We consider three
cases based on the placement of the sink vertex $t$.

\textbf{Case 1: }$t$ is mapped to a point of $C_i$
(Figure~\ref{fig:two_u}.a). It is easy to see that  arc $(s,u_k)$
crosses  arc $(u_j,t)$, a clear contradiction to the planarity of
the embedding.

\textbf{Case 2: } $t$ is mapped to $t(S)$
(Figure~\ref{fig:two_u}.b). Similar to the previous case since $C_i
\cup \{b(S),t(S)\}$ is a one-sided convex point set.

\textbf{Case 3: } $t$ is mapped to a point of
 $C_p,~p>i$, denote it by $p_t$ (Figure~\ref{fig:two_u}.c). By Statement~\ref{stat:prop_pointset} $C_i
\cup \{b(S),p_t\}$ is a convex point set and  points $p_t$, $b(S)$
are consecutive points of $C_i \cup \{b(S),p_t\}$. Hence, arc
$(s,u_k)$ crosses  arc $(u_j,t)$, a contradiction.  \qed

\begin{figure}[tb]
\centering
    \includegraphics[width=0.7\textwidth]{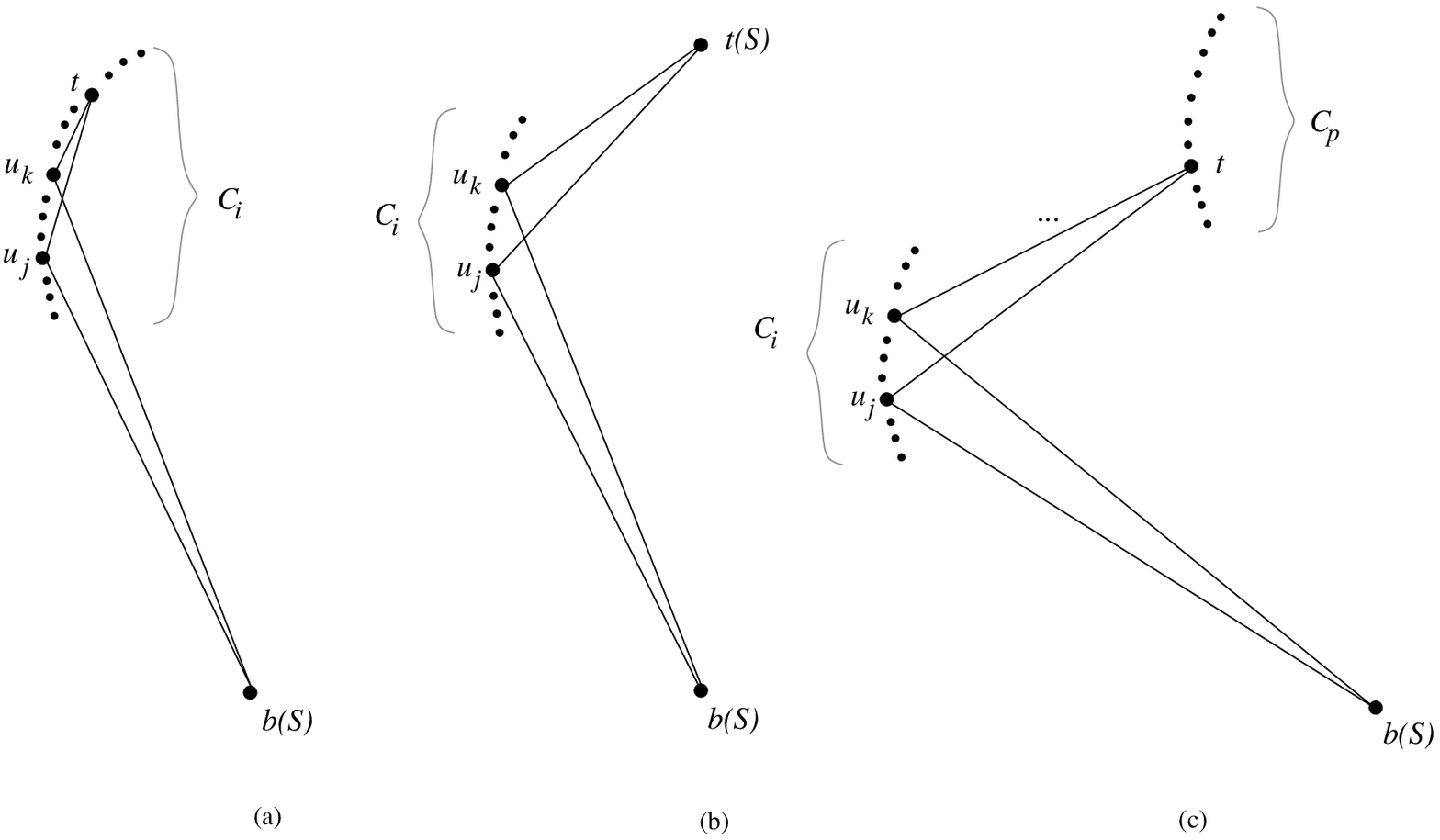}
    \caption{Mappings used in the proof of  Statement~\ref{statement:one_u}}
    \label{fig:two_u}
\end{figure}

\smallskip

By Statement~\ref{statement:one_u}, we have that each $C_i, i=1
\ldots m,$ contains exactly one vertex from set $\{ u_1, \ldots u_m
\}$. W.l.o.g., we assume that $u_i$ is mapped to a point of $C_i$.

\begin{statement}
\label{statement:top_vertex} In any UPSE of $G$ into $S$,  vertex
$t$ is mapped to either a point of $C_m$ or to $t(S)$.
\end{statement}
\emph{Proof of Statement~\ref{statement:top_vertex}:} $t$ has to be
mapped higher than any $u_i,~ i=1\ldots m,$ and hence higher than
$u_m$, which is mapped to a point of $C_m$. \qed

\begin{statement}
\label{statement:top_pointset} In any UPSE of $G$ into $S$, vertex
$u_i$ is mapped to $t(C_i)$, $1 \leq i \leq m-1$, moreover, there is
no arc $(v,w)$ so that $v$ is mapped to a point of $C_i$ and $w$ is
mapped to a point of $C_j$, $j>i$.
\end{statement}
\emph{Proof of Statement~\ref{statement:top_pointset}:} We prove
this statement by induction on $i,~i=1\ldots m-1$. For the basis,
assume that $u_1$ is  mapped to a point $p_1$ different from
$t(C_1)$ (see Figure~\ref{fig:highest_in_convex}.a). Let $p_t$ be
the point where vertex $t$ is mapped. By
Statement~\ref{statement:top_vertex},  $p_t$ can be either $t(S)$ or
a point of $C_m$. In both cases, point set $C_1 \cup \{b(S),p_t\}$
is a convex point set, due to the construction of the point set $S$
and the Statement~\ref{stat:prop_pointset}. Moreover, the points
$b(S)$ and $p_t$ are consecutive on the convex hull of point set
$C_1 \cup \{b(S),p_t\}$.

Denote by $p$ the point of $C_1$ that is exactly above the point
$p_1$. From Statement~\ref{statement:one_u}, we know that no $u_j$,
$j \neq 1$ is mapped to the point $p$. Due to
Statement~\ref{statement:top_vertex}, $t$ cannot be mapped to $p$.
Hence there is a path $P_k$, $1\leq k \leq 3m$, so that one of its
vertices is mapped to $p$. Call this vertex $u$. We now consider two
cases based on whether $u$ is the first vertex of $P_k$ of not.

\textbf{Case~1:} Assume that there is a vertex $v$ of $P_k$, such
that there is an arc $(v,u)$. Since the drawing of $S$ is upward,
$v$ is mapped to a point lower than $p$ and lower than $p_1$. Since
$C_1 \cup \{b(S),p_t\}$ is a convex point set, arc $(v,u)$ crosses
arc $(u_1,t)$. A clear contradiction.

\textbf{Case~2:} Let $u$ be the first vertex of $P_k$. Then, arc
$(s,u)$ crosses the arc $(u_1,t)$ since, again, $C_1 \cup
\{b(S),p_t\}$ is a convex point set, a contradiction.

So, we have that $u_1$ is mapped to $t(C_1)$, see
Figure~\ref{fig:highest_in_convex}.b. Observe now that any arc
$(v,w)$, such that $v$ is mapped to a point of $C_1$ and $w$ is
mapped to a point $x \in   C_2 \cup \dots \cup C_m \cup \{t(S)\}$
crosses arc $(s,u_1)$, since $C_1 \cup \{b(S),x\}$ is a convex point
set. So, the statement is true for $i=1$.

For the induction step, we assume that the statement is true for
$C_{g}$ and $u_{g}$, $g\leq i-1$, i.e. vertex $u_{g}$ is mapped to
$t(C_{g})$ and there is no arc connecting a point of $C_{g}$ to a
point of $C_k$, $k >g$ and this holds for any $g \leq i-1$. We now
show that it also holds for $C_{i}$ and $u_{i}$. Again, for the sake
of contradiction, assume that $u_i$ is mapped to a point $p_i$
different from $t(C_i)$ (see Figure~\ref{fig:highest_in_convex}.c).
\begin{figure}[tb]
\centering
    \includegraphics[width=\textwidth]{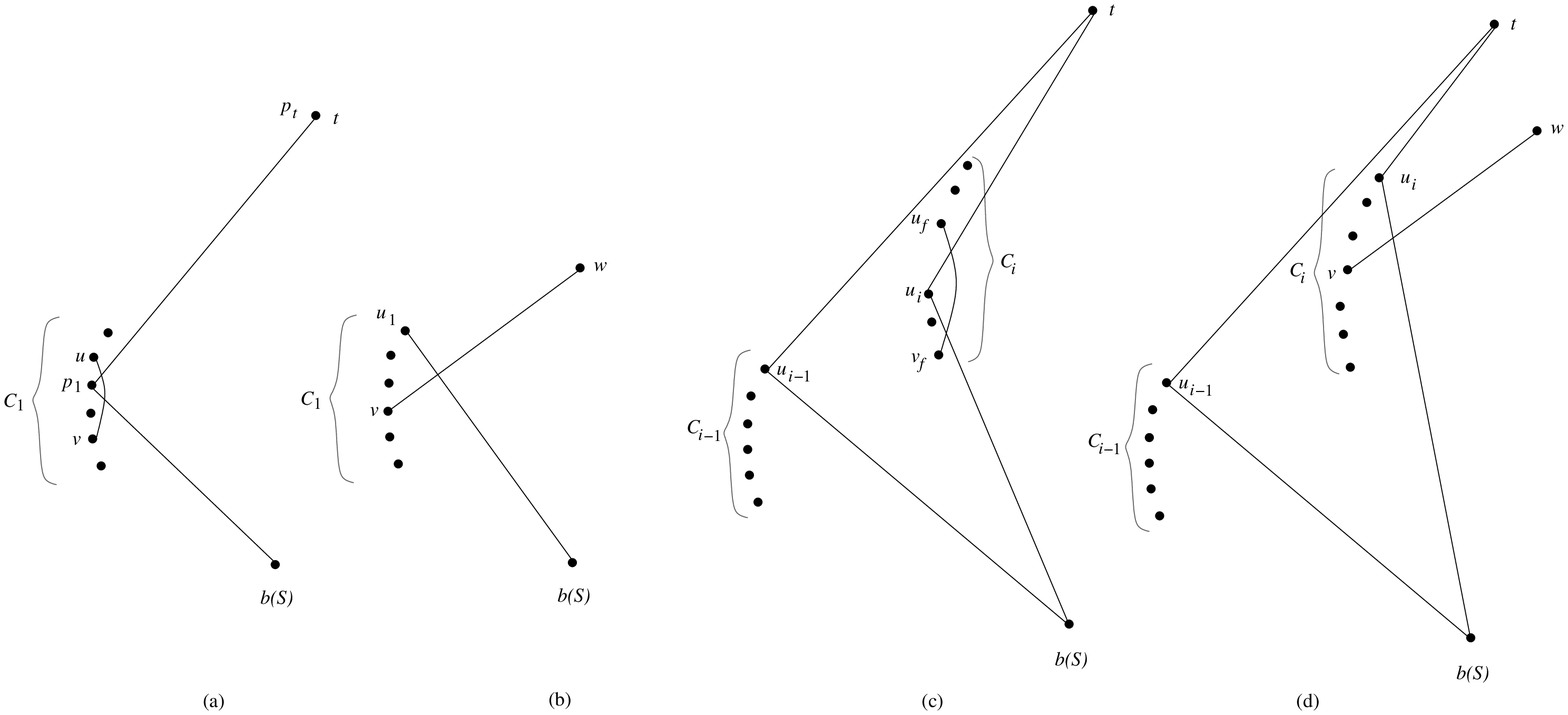}
    \caption{ Mappings used in the proof of Statement~\ref{statement:top_pointset}. }
    \label{fig:highest_in_convex}
\end{figure}

Denote by $q$ the point of $C_1$ that is exactly above point $p_i$.
From Statement~\ref{statement:one_u}, we know that no $u_l$, $l \neq
i$, is mapped to the point $q$. Due to
Statement~\ref{statement:top_vertex}, $t$ can not be mapped to $q$.
Hence, there is a path $P_f$, so that one of its vertices is mapped
to $q$. Call this vertex $u_f$. We now consider two cases based on
whether $u_f$ is the first vertex of $P_f$ of not.

\textbf{Case~1:} Assume that there is a vertex $v_f$ of $P_k$ such
that there is an arc $(v_f,u_f)$. By the induction hypothesis, we
know that $v_f$ is not mapped to any $C_l$, $l < i$. Then, since the
drawing of $S$ is upward, $v_f$ is mapped to a point lower than $q$
and lower than $p_i$. Since $C_i \cup \{b(S),p_t\}$ is a convex
point set, arc $(v_f,u_f)$ crosses arc $(u_i,t)$. A clear
contradiction.

\textbf{Case~2:} Let $u_f$ be the first vertex of $P_k$. Then, arc
$(s,u_f)$ crosses the arc $(u_i,t)$ since, again, $C_i \cup
\{b(S),p_t\}$ is a convex point set, a contradiction.

So, we have shown that $u_i$ is mapped to $t(C_i)$, see
Figure~\ref{fig:highest_in_convex}.d. Observe now that, any arc
$(v,w)$, such that $v$ is mapped to a point of $C_i$ and $w$ is
mapped to a point $x \in   C_{i+1} \cup \dots \cup C_m \cup
\{t(S)\}$ crosses arc $(s,u_i)$, since $C_i \cup \{b(S),x\}$ is a
convex point set. So, the statement holds for $i$. \qed

A trivial corollary of the previous statement is the following:

\begin{statement}
\label{statement:no_connection} In any UPSE of $G$ into $S$, any
directed path $P_j$ of $G$ originating at $s$, $j\in
\{1,\dots,3m\}$,
 has to be drawn entirely in $C_i$, for $i\in \{1,\dots,m\}$. \qed
\end{statement}

The following statement completes the proof of the theorem.

\begin{figure}[tb]
\centering
    \includegraphics[width=\textwidth]{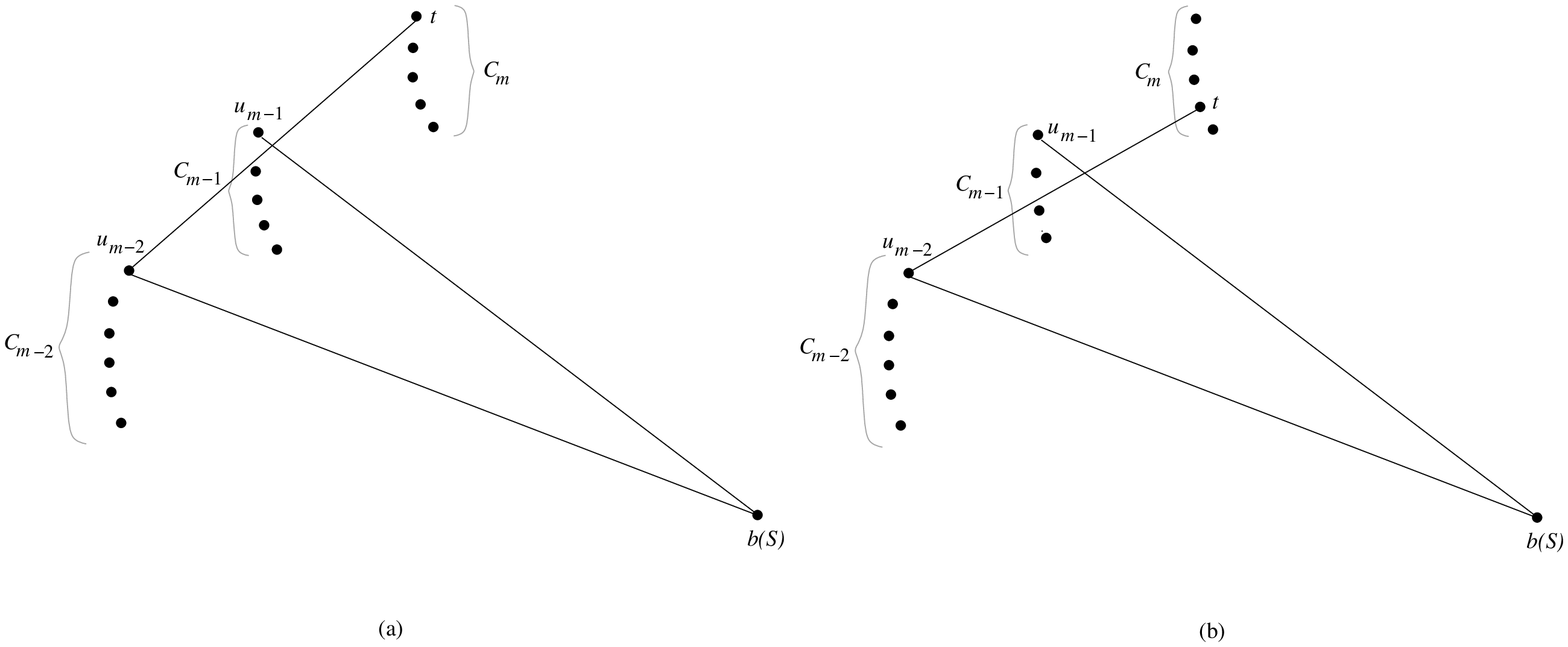}
    \caption{(a-b) Mappings used in the proof of Statement~\ref{statement:top_on_top}.}
    \label{fig:top_to_top}
\end{figure}

\begin{statement}
\label{statement:top_on_top} In any UPSE of $G$ into $S$,  vertex
$t$ is mapped to point $t(S)$.
\end{statement}
\emph{Proof of Statement~\ref{statement:top_on_top}:} For the sake
of contradiction, assume that  $t$ is not mapped to $t(S)$. By
Statement~\ref{statement:top_vertex} we know that $t$ has to be
mapped to a point in $C_m$. Assume first that $t$ is mapped to point
$t(C_m)$ (see Figure~\ref{fig:top_to_top}.a). Recall that $u_{m-2}$
and  $u_{m-1}$ are mapped to $t(C_{m-2})$ and $t(C_{m-1})$,
respectively, and that $\{t(C_i):i=1\ldots m\}$ is a left-heavy
convex point set. Hence,  points
$\{t(C_{m-2}),t(C_{m-1}),t(C_m),b(S)\}$ form a convex point set. It
follows that   segments $(t(C_{m-2}),t(C_m))$ and
$(t(C_{m-1}),b(S))$ cross each other, i.e.  edges $(s,u_{m-1})$ and
$(u_{m-2},t)$ cross, contradicting the planarity of the drawing.

Consider now the case where  $t$ is mapped to a point of $C_m$, say
$p$,   different from $t(C_m)$ (see Figure~\ref{fig:top_to_top}.b).
Since point $p$ does not lie in  triangle
$t(C_{m-2}),t(C_{m-1}),b(S)$ and  point $t(C_{m-1})$ does not lie in
 triangle $t(C_{m-2}),p,b(S)$,  points
$\{t(C_{m-2}),t(C_{m-1}),p,b(S)\}$ form a convex point set. Hence,
 segments $(t(C_{m-2}),p)$ and $(t(C_{m-1}),b(S))$ cross each
other, i.e.  edges $(s,u_{m-1})$ and $(u_{m-2},t)$ cross; a clear
contradiction.  \qed

Let us now combine the above statements in order to derive a
solution for the 3-Partition problem when we are given an UPSE of
$G$ into $S$. By Statement~\ref{statement:bottom} and
Statement~\ref{statement:top_on_top},  vertices $s$ and $t$ are
mapped to $b(S)$ and $t(S)$, respectively. By
Statement~\ref{statement:one_u}, for each $i=1 \ldots m$,  point set
$C_i$ contains exactly one vertex from $\{ u_1, \ldots, u_m \}$ ,
say $u_i$ and, hence,  the remaining  points of $C_i$ are occupied
by the vertices of some paths $P_i^1,P_i^2,\dots,P_i^c$. By
Statement~\ref{statement:no_connection}, $P_i^1,P_i^2,\dots,P_i^c$
are mapped entirely to the points of $C_i$. Since $C_i$ has $B+1$
points, the highest of which is occupied by  $u_i$, we infer that
$P_i^1,P_i^2,\dots,P_i^c$ contain exactly $B$ vertices. We set
$A_i=\{a_i^1,a_i^2,\dots,a_i^c\}$, where $a_i^j$ is the size of path
$P_i^j,~1 \leq j \leq c$. Since $\frac{B}{4}<a_i^j<\frac{B}{2}$ we
infer that $c=3$. The subsets $A_i$ are disjoint and their union
produces $A$.

Finally, we note that $G$ has a single source $s$ and the longest
simple cycle of $G$ has length $4$, moreover the point set $S$ is an
$m$-convex point set for some $m>1$. This completes the proof.\qed

\end{proof}

\section{Open Problems}
\label{se:conclusions} In this paper, we continued the study of the
upward point-set embeddability problem, initiated
in~\cite{AngeliniFGKMS10,BinucciGDEFKL10,GiordanoLMS07}. We showed
that the problem is NP-complete, even if some restrictions are posed
on the digraph and the point set. We also extended the positive and
the negative results presented
in~\cite{AngeliniFGKMS10,BinucciGDEFKL10} by resolving the problem
for the class of $k$-switch trees, $k\in \mathbb{N}$. The partial
results on the directed trees presented
in~\cite{AngeliniFGKMS10,BinucciGDEFKL10} and in the present work,
may be extended in two ways: $(i)$ by presenting the time complexity
of the problem of testing whether a given directed tree admits an
upward planar straight-line embedding (UPSE) to a given
general/convex point set and $(ii)$ by presenting another classes of
trees, that admit/do not admit an UPSE to a given general/convex
point set. It would be also interesting to know whether there exists
a class of upward planar digraphs $\mathcal{D}$ for which the
decision problem whether a digraph $D \in \mathcal{D}$ admits an
UPSE into a given point set $P$ remains NP-complete even for a
convex point set $P$.

\bibliography{bibliography}
\bibliographystyle{plain}

\end{document}